\documentclass[12pt]{article}

\newcommand{\be}{\begin{equation}}
\newcommand{\ee}{\end{equation}}
\newcommand{\bea}{\begin{eqnarray}}
\newcommand{\eea}{\end{eqnarray}}
\newcommand{\br}{\hskip .25cm/\hskip -.25cm}
\newcommand{\hf}{\frac{1}{2}}
\newcommand{\nn}{\nonumber\\}

\newcommand{\ol}{\overline}
                                                                                                                     
\begin{document}
                                                                                                                     
\begin{center}
{\Large{\bf An alternative to exact renormalization equations}}

\vspace{.5cm}

{\bf Jean Alexandre}\\ 
Physics Department, King's College,\\ 
London WC2R 2LS, UK\\
jean.alexandre@kcl.ac.uk

\vspace{.5cm}

Based on a talk given in the\\ 
"Workshop on Recent Advances in Particle Physics and Cosmology",\\
Thessaloniki 2005.

\vspace{1cm}
{\bf Abstract}

\end{center}

An alternative point of view to exact renormalization equations is discussed, where 
quantum fluctuations of a theory are controlled by the bare mass of a particle. 
The procedure is based on an exact evolution equation for the effective action, 
and recovers usual renormalization results.

\vspace{1cm}

\section{Introduction}

In the framework of exact renormalization equations in Quantum Field Theory, 
an intuitive blocking procedure is provided by the Wegner-Houghton equation
\cite{wegnerhoughton}, where Fourier modes of a scalar theory are infinitesimally eliminated  
with the use of a sharp cut off, from the scale $k$ to $k-\delta k$. In the limit where
$\delta k\to 0$, the authors find an exact equation describing the evolution with $k$ of the 
effective theory describing the dynamics of Fourier modes with momentum less than $k$. 
This procedure, however, suffers from a drawback: if the infrared field
is not uniform, the limit $\delta k\to 0$ generates singularities due to the 
non differentiability of the sharp cut off. 

A solution is given by the use of a smooth cut off with typical momentum $k$ \cite{polchinski, others}.
This procedure, though, is based on a cut off function introduced by hand. 
It has been argued however that the action obtained in the limit
where $k\to 0$, the quantum theory is independent of the choice of this cut off function. 
For a review, see \cite{wetterich} and references therein.   

The sharp cut off procedure suffers from another problem: it is not gauge invariant.
This problem is dealt with in the framework of the smooth cut off procedure,
where gauge invariance is recovered for the proper graphs generator functional.

We present here an alternative approach to the construction of quantum theories, also based on
an exact functional method. The idea is to select degrees of freedom by the amplitude of
their quantum fluctuations, instead of their Fourier modes. As a consequence, problems related to 
a running cut off do not occur here.
Controlling quantum fluctuations is achieved 
with the help of a mass in the bare theory: if this bare mass is large compared to any other mass scales,
quantum fluctuations are frozen and the quantum theory is the classical one. As this mass
decreases, quantum fluctuations gradually appear and the theory gets dressed. The interesting point
is that one can obtain an exact evolution equation for the proper graphs generator functional with the 
amplitude of the bare mass.

Section 2 describes this method in the case of a scalar theory and shows the equivalence with
usual renormalization methods.

Section 3 applies these ideas to QED, where the flows in the bare mass exhibit the analogous behaviour 
to the Landau pole, if we approximate the equations to one loop. To proceed beyond one loop, 
the equations are solved without approximation and it is found that this Landau pole behaviour vanishes.

Section 4 deals with other models where this functional method is applied: QED and
Wess-Zumino model in 2+1 dimensions, and Liouville theory.

Finally, a review of the different functional methods given in this introduction 
can be found in \cite{janos}, and a detailed description of this alternative approach
to renormalization is given in \cite{janoskornel}.

\eject

\section{General procedure}

We describe here the general procedure with a scalar model. The bare theory is
\be
S_\lambda[\Phi]=
\int d^4x\left\{\frac{1}{2}\partial_\mu\Phi\partial^\mu\Phi-\lambda\frac{m_0^2}{2}\Phi^2
-\frac{g}{24}\Phi^4\right\},
\ee
where the dimensionless parameter $\lambda$ controls the amplitude of the bare mass. To obtain the
effective action (the proper graphs generator functional), one proceeds in the usual way, which is 
reminded here.

The connected graphs generator functional is
\be
\exp\{iW_\lambda[j]\}=\int{\cal D}[\Phi]\exp\left\{iS_\lambda[\Phi]+i\int d^4x~j(x)\Phi(x)\right\},
\ee
and the quantum field is defined as 
\be
\phi(x)=-i\frac{\delta W_\lambda}{\delta j(x)}.
\ee
Inverting the relation between $\phi$ and $j$, 
one then performs the Legendre transform of $W_\lambda$ with respect to the source $j$, which
leads to the effective action $\Gamma_\lambda$:
\be
\Gamma_\lambda[\phi]=W_\lambda[j]-\int d^4x~j(x)\phi(x).
\ee
One should keep in mind here that the independent variables of $\Gamma$ are 
$\phi$ and $\lambda$ and that $j$ has to be seen a function of these variables.

The idea is to start from a large value $\lambda>>1$, which has the effect to freeze 
quantum fluctuations. In this situation, one expects that $\Gamma_\lambda[\phi]=S_\lambda[\phi]$. As 
$\lambda$ decreases, quantum fluctuations gradually appear and the theory gets dressed. Finally,
when $\lambda=1$, one reaches the full quantum theory. We will see that this procedure indeed recovers
the well know results.

The interesting point is that one can obtain for $\Gamma_\lambda$ an 
exact evolution equation with $\lambda$, which is \cite{scalar}:
\be\label{evolscal}
\partial_\lambda\Gamma_\lambda[\phi]
=\frac{m_0^2}{2}\mbox{Tr}\left\{-\phi^2+i\left(\frac{\delta^2\Gamma_\lambda}
{\delta\phi\delta\phi}\right)^{-1}\right\},
\ee
where Tr denotes the trace of the operator inside the brackets. Note the similarity with usual
renormalization equations, but without mentioning any running cut off.

The next step is to assume a functional dependence of $\Gamma_\lambda$ with $\phi$. Let us consider for this
the following gradient expansion:
\be\label{gradexp}
\Gamma_\lambda[\phi]=
\int d^4x\left\{\frac{1}{2}Z_\lambda(\phi)\partial_\mu\phi\partial^\mu\phi-U_\lambda(\phi)\right\}.
\ee
Plugging this assumption into the evolution equation (\ref{evolscal}) leads to differential equations
for $Z_\lambda$ and $U_\lambda$, which are of the form:
\bea\label{evolscalexp}
\partial_\lambda Z_\lambda(\phi)&=&\hbar F_Z\Big(\lambda,Z_\lambda(\phi),U_\lambda(\phi)\Big)\nn
\partial_\lambda U_\lambda(\phi)&=&\hbar F_U\Big(\lambda,Z_\lambda(\phi),U_\lambda(\phi)\Big),
\eea
where the functions $F_Z,F_U$ are regulated by a cut off and $\hbar$ was restored for clarity.
Note that these equations look like one-loop equations, but they are not since the right-hand
side contains the the partially dressed functions $(Z_\lambda,U_\lambda)$. 

So as to find the full quantum theory,  
these equations have to be integrated from $\lambda=\infty$ to $\lambda=1$. To look at the consistency at 
one-loop, one can replace in the right hand side of Eqs.(\ref{evolscalexp})
the functions $(Z,U)$ by their bare values and then integrate over $\lambda$:
\bea\label{equaZU}
Z_1(\phi)&=&Z_{bare}(\phi)
+\hbar\int_\infty^1 d\lambda~F_Z\Big(\lambda,Z_{bare}(\phi),U_{bare}(\phi)\Big)+{\cal O}(\hbar^2)\nn
U_1(\phi)&=&U_{bare}(\phi)
+\hbar\int_\infty^1 d\lambda~F_U\Big(\lambda,Z_{bare}(\phi),U_{bare}(\phi)\Big)+{\cal O}(\hbar^2),
\eea
where 
\bea
Z_{bare}(\phi)&=&1\nn
U_{bare}(\phi)&=&\lambda\frac{m_0^2}{2}\phi^2+\frac{g}{24}\phi^4.
\eea
It was indeed found in \cite{scalar} that Eqs.(\ref{equaZU}) give the usual results.  
To see this, we give here the explicit equation for $U$ (in Euclidean space):
\bea\label{oneloopscal}
&&U_1(\phi)\\
&=&U_{bare}(\phi)+\frac{\hbar m_0^2}{2}\int_\infty^1 d\lambda\int\frac{d^4p}{(2\pi)^4}
\frac{1}{p^2+\lambda m_0^2+g\phi^2/2}+{\cal O}(\hbar^2)\nn
&=&U_{bare}(\phi)+\frac{\hbar}{2}\int\frac{d^4p}{(2\pi)^4}\ln\left(\frac{p^2+m_0^2+g\phi^2/2}
{p^2+m_0^2}\right)+\mbox{C}+{\cal O}(\hbar^2),\nonumber
\eea 
where C is an infinite field-independent constant and 
the integration over Fourier modes has to be regularized. The expression (\ref{oneloopscal})
is the well-known one-loop potential.
In addition, it was shown that in the ultraviolet regime, the flows in $\lambda$ reproduce the well-known
one-loop renormalization flows. 
As a result, the method described here indeed provides a renormalization procedure.

\section{Quantum Electrodynamics}

We wish to apply the same ideas to QED and to control quantum fluctuations with the fermion mass.
The starting point is the following bare action, in dimension $4-\varepsilon$: 
\be
S_\lambda=\int d^{4-\varepsilon}x~A_\mu\frac{T^{\mu\nu}+\alpha L^{\mu\nu}}{e^2\mu^\varepsilon}A_\nu
+\ol\psi(i\br D-\lambda m_0)\psi,\nonumber
\ee
where $T^{\mu\nu}$ and $L^{\mu\nu}$ are the transverse and longitudinal projectors respectively and 
$\alpha$ is a usual gauge fixing parameter. The exact evolution equation that is obtained 
for the effective action is \cite{qed4}:
\be
\partial_\lambda\Gamma_\lambda
=m_0\mbox{Tr}\left\{-\ol\psi\psi+\left(\frac{\delta^2\Gamma_\lambda}
{\delta\ol\psi\delta\psi}\right)^{-1}\right\}.
\ee
To solve this equation, we consider the following ansatz:
\be\label{ansatzqed}
\Gamma_\lambda=\int d^{4-\varepsilon}x~A_\mu
\frac{\beta_T(\lambda)T^{\mu\nu}+\alpha\beta_L(\lambda)L^{\mu\nu}}{e^2\mu^\varepsilon}A_\nu
+\ol\psi\Big(z(\lambda)i\br D-m(\lambda)\Big)\psi,
\ee
i.e. we do not take into account additional operators, but allow the different coefficients 
of the original action to evolve with $\lambda$. The evolution equations that are obtained read then
\bea\label{equaqed4}
\frac{d\beta_L}{d\lambda}&=&0\nn
\frac{d\beta_T}{d\lambda}&=&
\frac{e^2\mu^\varepsilon}{6\pi^2}\frac{m_0}{m(\lambda)}\nn
\frac{dz}{d\lambda}&=&\frac{e^2\mu^\varepsilon}{8\pi^2\alpha}
z(\lambda)\frac{m_0}{m(\lambda)}\nn
\frac{dm}{d\lambda}&=&-\frac{m_0}{8\pi^2}\frac{e^2\mu^\varepsilon}{\varepsilon}
\left(\frac{3}{\beta_T(\lambda)}+\frac{1}{\alpha}\right)
\eea
Note the following points:

\begin{itemize}
\item the non physical longitudinal part of the photon propagator does not evolve with $\lambda$; 
\item as expected, the photon does not acquire a mass in the evolution with $\lambda$;
\item the equation for the evolution of the fermion mass is the only one that needs to be regularized;
\item the wave function renormalization $z$ keeps its initial value $z=1$ in the Landau gauge $\alpha=\infty$,
as in a usual renormalization procedure.
\end{itemize}

One can check the consistency with known one-loop result, as was done in the scalar case, and in particular 
for the evolution of $\beta_T$: if the fermion running mass is replaced by its bare value $\lambda m_0$ 
in the equation for $\beta_T$, we find after integration over $\lambda$ (the limit $\varepsilon\to 0$
can be taken):
\be
\beta_T(\lambda)=1+\frac{e^2}{6\pi^2}\ln\left(\frac{\lambda}{\lambda_0}\right),
\ee
where $\lambda_0>>1$ such that $\beta_T(\lambda_0)=1$. Therefore $\beta_T$ vanishes for the value
\be
\frac{\lambda_0}{\lambda}=\exp\left(\frac{6\pi^2}{e^2}\right),
\ee
which is the well know one-loop value for the Landau pole if we identify $\lambda_0/\lambda$ with 
a ratio of momenta. If the infrared theory is fixed (at $\lambda=1$), we find the Landau pole-like 
singularity occurring for the ultraviolet value $\lambda_0$. The $\lambda$-flow thus reproduces the expected 
one-loop singularity in the momentum-flow of the vacuum polarization . 

To go beyond one-loop, one can integrate eqs. (\ref{equaqed4}) simultaneously to obtain 
\be
\beta_T(\lambda)=\left(\frac{m(\lambda_0)}{m(\lambda)}\right)^{4\varepsilon/9}
\exp\left\{\frac{1-\beta_T(\lambda)}{3}\left(\frac{1}{\alpha}-
\frac{8\pi^2\varepsilon}{e^2\mu^\varepsilon}\right)\right\},
\ee
where the dimensional regularization is kept. 
We see here that $\beta_T$ does not vanish anymore: the Landau pole disappears if we take into account 
the evolution of the fermion mass. This result is consistent with \cite{landaupole} 
where it is shown that the Landau pole does not occur if chiral symmetry breaking is allowed, which is 
another way to consider the evolution of the fermion mass with quantum fluctuations.

\section{Other examples}

\subsection{Parity-conserving $QED_3$}

The motivation to consider $QED_3$ is the study of dynamical mass generation, usually done with 
Dyson-Schwinger method \cite{qed2+1}. The latter does not allow the progressive appearance of 
quantum fluctuations but directly "jumps" to the effective theory. On the contrary, the method 
exposed here sets the quantum theory in a progressive manner and thus is expected to be 
more accurate.

This work is done in the following framework:

\begin{itemize}
\item we consider an even number (2N) of fermion flavours, with opposite chiralities, 
so as to cancel the generation of Chern-Simons terms; 
\item no regularization is needed: UV divergences are absent due to low dimensionality,
and IR divergence are also absent due to dynamical mass generation;
\item for the functional form of $\Gamma_\lambda$, we consider the same ansatz as in 3+1 dimensions, 
given by eq.(\ref{ansatzqed}). 
\end{itemize}

The evolution equation is similar to the one obtained in 3+1 dimensions and
to study dynamical mass generation, one integrates this evolution equation from 
$\lambda=\infty$ to \underline{$\lambda=0$} since we are interested in a vanishing bare mass. As a result, 
the following relation between the dynamical mass $m_{dyn}$ and the renormalized coupling $g_R$ is found
\cite{qed3}:
\be
g_R=g_B\left(1+\frac{3N+4}{18\pi}\frac{g_B^2}{m_{dyn}}\right)^{-2/(3N+4)},
\ee
where $g_B$ is the (dimensionfull) bare coupling. This non perturbative result is a self consistent relation
between dressed parameters and shows that, in this approximation for the effective action, an 
interacting theory ($g_R\ne 0$) necessarily generates a dynamical mass ($m_{dyn}\ne 0$).

This study is being extended to a momentum dependent fermion self energy to
study the influence of the number of flavours on the dynamical mass generation.

\subsection{Planar Wess-Zumino model}

$N=1$ supersymmetry in 2+1 dimensions does not have non-renormalization theorems:  
the odd coordinate of superspace is real, and non-renormalization theorems are based on the
analyticity of quantum corrections in the complex Grassmann coordinate.
Due to supersymmetry though, we can expect interesting properties of the quantum theory. 

The bare action for the Wess-Zumino model we consider is
\bea
S_\lambda=\int d^5z\left\{\hf QD^2Q+\lambda\frac{m_0}{2}Q^2+\frac{g}{24}Q^4\right\},
\eea 
where $Q$ is a real superfield and $z=(x,\theta)$ is the superspace coordinate (where $\theta$ is real).
The interaction $Q^4$ leads to an on shell $\phi^6$ scalar potential, such that $g$ is a dimensionless 
coupling constant. The evolution equation is \cite{WZ}
\be
\partial_\lambda\Gamma_\lambda[Q]=\frac{m_0}{2}\mbox{Tr}\left\{Q^2+
\left(\frac{\delta^2\Gamma_\lambda}{\delta Q\delta Q}\right)^{-1}\right\}.
\ee
We consider the local potential approximation:
\be\label{approxWZ}
\Gamma_\lambda=\int d^5z\left\{\hf QD^2Q+U_\lambda(Q)\right\},
\ee
where we allow the evolution of the potential only, and keep the kinetic term fixed.
With this ansatz, the evolution equation happens to be linear in the potential, as a 
consequence of supersymmetry. It is thus possible to derive the exact solution:
\be
U_\lambda(Q)=\lambda\frac{m_0}{2}\left(1+\frac{g}{8\pi}\right)Q^2+\frac{g}{24}Q^4,
\ee
and we see that there is no renormalization of the interaction; only the mass term gets dressed.
It should not be forgotten that this result is valid in the approximation (\ref{approxWZ}) only and that 
further studies should be made, that incorporate higher order terms in the gradient expansion.

\subsection{Liouville theory}

So as to restore Weyle invariance in the quantization of bosonic strings propagating
in a target space with $d\ne 26$, one has to introduce an additional
degree of freedom: the Liouville field. This scalar field lives on the string world sheet and has 
dynamics described by the following bare action:   
\be
S_\lambda=\int d^2\xi\sqrt{g}\left\{\hf g_{ab}\partial^a\phi\partial^b\phi
+R\phi+\lambda m^2e^{\phi}\right\},
\ee
where $g_{ab}$ is the world sheet metric, $g$ is its determinant and $R$ is the associated curvature scalar. 
For a review on the Liouville formalism, see \cite{ginsparg}, and 
for a review on phenomenological implications, see \cite{mavromatos}.

From the renormalization point of view, an interesting point is that $\phi$ is dimensionless,
such that any power of $\phi$ is a classically marginal operator. So is the exponential potential, which
has the following symmetry: a change in $\lambda$ is equivalent to a translation in the field.
As a consequence, one obtains for the effective action the following equation \cite{liouville}:
\be\label{linear}
\lambda\partial_\lambda\Gamma_\lambda+4\pi\chi
=\int d^2\xi\frac{\delta\Gamma_\lambda[\phi]}{\delta\phi(\xi)},
\ee
where $\chi$ is the Euler characteristic of the world sheet. 
Unlike the previous models, the idea to solve this equation would be to integrate it from $\lambda=0$, where
no interaction occurs and thus the effective action is the classical one. But
note that this equation is linear in $\Gamma_\lambda$: one cannot consider the initial value
$\Gamma_0=S_0$, otherwise no quantum correction occurs when $\lambda>0$. This is another way to 
recover the well-known non perturbative aspect of Liouville theory: the quantum theory cannot
be obtained by switching on gradually the mass parameter $m^2$. This point was already discussed
using usual exact renormalization equations \cite{wliouville}.

A first consequence of this linearity is a set of sum rules for the proper functions, defined as 
functional derivatives of $\Gamma_\lambda$ with respect to $\phi$, for vanishing field:
\be
\lambda\partial_\lambda G^{(n)}_\lambda(\xi_1,...,\xi_n)
=\int d^2\xi~G^{(n+1)}_\lambda(\xi,\xi_1,...,\xi_n).
\ee
Sum rules for exponential operators were also obtained in \cite{pawlowski}.

Another consequence of the linearity of eq.(\ref{linear}) can be seen by taking successive derivatives 
with respect to $\lambda$ and then make a ressumation over the derivatives with respect to the field, 
to obtain \cite{liouville}
\be
\Gamma_\lambda[\phi]=\Gamma_1[\phi+\ln\lambda]-4\pi\chi\ln\lambda
\ee
Therefore, as the bare action, the effective action $\Gamma_\lambda$ depends on the combination $\phi+\ln\lambda$.

\section{Conclusion}

To conclude, we can stress the following characteristics of the method presented here:

\begin{itemize}
\item it provides another way to generate a quantum theory;
\item it is based on an exact, non-perturbative, functional equation;
\item it is an alternative to coarse graining and reproduces the well known one-loop renormalization flows,
without assuming any cut off function.
\end{itemize}

The study of non-Abelian theories is possible with this method since the derivative with respect to
$\lambda$ corresponds to the insertion of a fermion propagator, such that even gluon loops can be controlled by
the bare fermion mass.

To study theories where no mass term is present, one can in principle follow the evolution of the
effective action with a coupling constant. The latter being in factor of a cubic or quartic operator though,
the exact equation satisfied by the effective action is more difficult to handle. A solution 
to this problem is left for future investigations, using for example a composite operator formalism.

\end{document}